\documentclass[prl,aps,twocolumn,showpacs]{revtex4-2}
\usepackage{graphicx,epstopdf}
\usepackage{dcolumn}
\usepackage{bm}
\usepackage{color}
\usepackage{xcolor}
\usepackage{float}
\usepackage{amssymb,amsmath}
\begin{document}

\title{Exact current blockade maps of dsDNA bound motifs driven through a solitary nanopore using electrokinetic Brownian dynamics}

\author{Swarnadeep Seth}
\author{Aniket Bhattacharya}
\altaffiliation[]
{Author to whom the correspondence should be addressed}
{}
\email{Aniket.Bhattacharya@ucf.edu}
\affiliation{Department of Physics, University of Central Florida, Orlando, Florida 32816-2385, USA}
\date{\today}
\begin{abstract}
  We report current blockade (CB) characteristics of molecular motifs residing on a model dsDNA using electrokinetic Brownian dynamics (EKBD) and study the role of the valence of the counter-ions as the dsDNA translocates through a solitary nanopore (NP) driven by an electric field. We explicitly incorporate all the charges on the DNA backbone, co- and counter-ions, and investigate CB characteristics of two charged sidechain motifs exactly. Our simulation brings out the details of binding and unbinding of the counter-ions and the time dependent counter-ion condensation on the translocating DNA  for mono- and di-valent salt conditions. An important and less intuitive finding is that the drop in the conventional (positive) current through the pore is due to the condensation of the counter-ions on the translocating DNA and not so much due to drop in the co-ions passing through the pore. This finding aligns with previous studies conducted by Tanaka {\em et al.} [Phys. Rev. Lett. {\bf 94}, 148103 (2005)], Cui [J. Phys. Chem B {\bf 114}, 2015 (2010)], and Holm {\em et al.} [Phys. Rev. Lett. {\bf 112}, 018101 (2014)]. We further find that this condensation is larger for the divalent ions leading to a slowing down of the translocation speed and yielding a longer dwell time for the motifs. Finally, we use the exact CB characteristics from the EKBD simulation to reconstruct the same CB characteristics using a volumetric ansatz on the segment inside and in the vicinity of the pore using on the ordinary BD model without the explicit presence of co- and counter-ions. Refinement of this ansatz will allow us to obtain the CB characteristics for longer genome fragments using low-cost ordinary BD simulation. 
\end{abstract}
\maketitle
Several decades of research have witnessed the growth and exploration of the potential of nanopore sensing~\cite{Review0}-\cite{Deamer1999} in bimolecular analysis and its applications in various fields~\cite{Wang2021,Wei,Aksimentiev2017,Keyser2018,Lopez}. Measurements of ionic current flowing through nanopores in biological or solid-state membranes can provide detailed information about both the properties of the nanopores and the solutes that pass through them~\cite{Deamer1996}. The nanopore sensing approach is a powerful tool that has been used to study a wide range of biomolecular systems, from ions to viruses~\cite{Roozbahani,Akhtarian}. It has the potential to revolutionize protein characterization and sequencing technologies, molecular biomarker detection~\cite{Bhatti,Chen2023,Dunbar2022}, drug design~\cite{Jeong}, mass spectrometry~\cite{Zhang2020}, and general purpose analytics~\cite{Wang2020}. However, realizing these goals requires the development of custom nanopores that produce well-defined current modulations and simultaneous understanding of the current blockade characteristics in the presence of target analytes. \par
Nanopore sequencing and biomarker identification face the challenge of fast translocation speeds~\cite{Wanunu2015,Yuan2020}, resulting in short current blockades. Strongly driven translocations exhibit non-uniform DNA velocity due to tension propagation along the chain backbone~\cite{Ikonen2011,Seth2021,Keyser2021}, while weakly driven conditions have a lower capture probability~\cite{Seth_2022}. To address these issues, various experimental and computational methods have been proposed. These include narrowing the pore diameter, altering solvent conditions~\cite{Fologea2005}, incorporating mesh obstacles at the pore entrance~\cite{Yan}, and implementing salt~\cite{Wanunu2010,Bello2019} and viscosity~\cite{Tsutsui2022} gradients across the pore. Recent studies have shown promising results in slowing down the translocation speed through the alteration of salt concentration~\cite{Dekker2012} and ion valency~\cite{Hsiao}.  
\begin{figure*}[ht!]
\begin{center}
\includegraphics[width=0.95\textwidth]{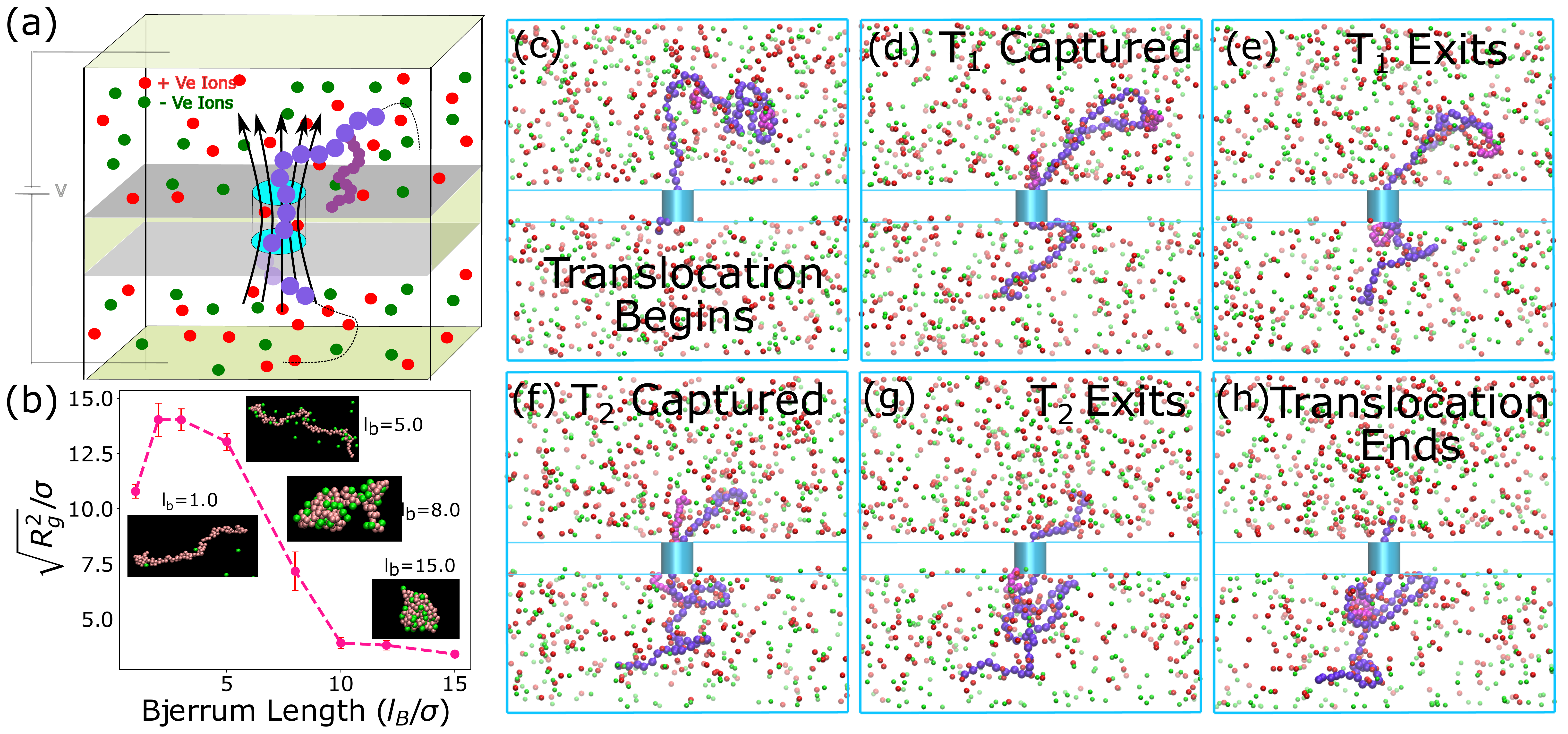}
\end{center}
\caption{\small (a) Schematics of electrokinetic translocation of a dsDNA through a solitary nanopore. The dsDNA is represented as a bead-spring polymer (violet) with a sidechain motif (magenta beads) attached to the backbone. The red/green beads represent monovalent positive/negative ions. The electric file (indicated by the arrow) due to a voltage bias $V$ across the nanopore facilitates translocation. (b) $R_g$ of a 128 bead long charged dsDNA in presence of 42 counter-ions as a function of the Bjerrum length as described in the text. The insets show the corresponding  snapshots of the counter-ion (in green) condensing on the DNA backbone (in pink) at approximate locations for the $l_B$ and eventual collapse of the chain. Snapshots of various stages of translocation of the motif attached to the dsDNA in presence of monovalent ions (c) marks the initiation; (d)-(e) and (f)-(g) depict the capture, translocation, and exit of the first ($T_1$) and the second $T_2$ motif respectively. Finally,  (h) marks the end of the translocation.}
\label{Model}
\end{figure*}
\par
Nanopore translocation experiments typically cover a wide time range, from microseconds to milliseconds. However, in coarse-grained (CG) frameworks, both length and time scales are modified, resulting in faster dynamics that can span the experimental time ranges. To establish a connection and comparison between experimental findings and simulations, it is possible to utilize dimensionless quantities such as scaling relations and measure statistical properties. In a previous study, we demonstrated that the Péclet number can be employed to compare experimental translocation data with simulations across a broad range of nanopore systems, including the double nanopore platform~\cite{Seth_DNP,Reisner2020}. In the context of making simulation models of translocation dynamics, it is necessary to compare the ionic current data obtained from experiments. The current trace is believed to encompass the comprehensive description of the translocation process, including the crucial signatures of biomarker specificity that need to be deciphered. All atom simulations~\cite{Li-Current, Aksimentiev2010,Marino} provide reasonably accurate data to generate the exact current blockade maps. However, these models are more computationally intensive than the continuum models~\cite{Ghosal,Fyta2011,Fyta2019,Kavokine}, relying on solving the Poisson-Nernst-Planck (PNP) equations, but they can account for the effects of solvent on ionic transport. On the other hand the continuum description lacks the specificity of the DNA-ion interactions. To reduce the computation complexities but to understand the overall dynamics and current blockade features, we have considered a CG model of a dsDNA with the explicit consideration of co-ions and counter-ions with mono- and di-valent salts (Fig.~\ref{Model}(a)). This approach enables a meaningful comparison and interpretation of results between experiments and simulations.
\par
Hence, the CG model strikes a balance with computational efficiency albeit embracing the essential characteristics of the system. In this article, we utilize Electrokinetic Brownian dynamics (EKBD) simulations to gain insights into the microscopic mechanisms of ion migration through a nanopore in the presence of double-stranded DNA (dsDNA) decorated with two sidechain motifs. Our main focus is on achieving an accurate yet fast {\em in-silico} construction of current blockades using the BD simulation. By employing our BD model, we are able to investigate the adsorption of counter-ions under low ionic concentration, considering both mono- and divalent salts. We introduce a volumetric occupation-based model and construct a current diagram during the translocation. To validate the reliability of our constructed current signatures, we compare them with the exact current obtained from ion migration across the pore. Furthermore, we conduct a comprehensive analysis and discussion to explore the impact of salt valency on the translocation speed of double-stranded DNA (dsDNA) as well as the dwell time of the two charged motifs. This investigation is particularly relevant for providing valuable insights that contribute to enhancing the accuracy and interpretability of current-based readouts in single nanopore experiments.
\par
{\bf $\bullet$ The Model and the method:}~
The EKBD simulation is performed on a CG model consisting of 64 beads long ($N=64$) dsDNA of diameter $\sigma_{bead}=2.0\sigma$ ($L=N\sigma_{bead}$) and of mass $m_{bead}=2m_{ion}$, where $m_{ion}$ is the mass of the co-ion or counter-ion chosen to be our unit of mass.  Two 12 units long sidechain motifs ($L_{motif}=12\sigma_{bead}$), each bead carrying unit charge (shown in magenta circles in Fig.~{Model}(a)) are attached along the dsDNA backbone. The diameter of the side chain monomers are $\sigma_{motif}=1.0\sigma$, half of the size of a dsDNA monomer to represent single-stranded DNA oligomers. To mimic the single nanopore experiments, we construct a solitary nanopore at the center of the simulation box (see Fig.~{Model}(a)) of length $L_{box}=64\sigma$ by removing a beads from the rigid wall of immobile LJ particles. The nanopore of width and thickness $4\sigma$ allows passage of the dsDNA with sidechains on its backbone and both types of ions. 
\par
In the coarse graining procedure, we fix the smallest length scale unit to be $\sigma$=1 nm. This translates the DNA double helix diameter to be 2 nm. We populate the simulation box of with 1000 mono-valent ions. The dsDNA is then released at the nanopore orifice.  In the hydrated condition, a typical ion's diameter is 0.5 - 0.7 nm \cite{Junior}, which is in the ball-park of $\sigma_{ion}=1.0\sigma$ (1 nm) considered in our simulation. In addition, we assume a strong ionization condition. Hence, each of the dsDNA-bead and  two oligomers-beads each release one counter-ion into the solution. We adjust the counter-ions and co-ions numbers accordingly, so that the system remains charge neutral with $N_{bead}\cdot q_{bead} + N_{motif}\cdot q_{motif} + N^{+ve}_{ion}\cdot q^{+ve}_{ion} + N^{-ve}_{ion}\cdot q^{-ve}_{ion} = 0$. Here $N_{bead}$, $N_{motif}$, $N^{+ve}_{ion}$, and  $N^{-ve}_{ion}$ represent the number of monomers in the dsDNA, motifs, counter-ions and co-ions are  respectively.  The corresponding charges are denoted by $q_{bead}$, $q_{motif}$, $q^{+ve}_{ion}$, and $q^{-ve}_{ion}$.
These parameters produce a 6.3 mM buffer that falls within experimental values (please see supplementary section).
\par
The electric field from the external voltage ($V$) (Fig.~\ref{Model}) produces a downward force of 2.4$k_BT/\sigma$ on a $-1.0e$ charge within the nanopore constriction that falls under the ``weekly-driven(WD)" regime~\cite{Sakaue,Hsiao} and comparable to the experimental voltage biases~\cite{Ito2016,Bello2019}. The simulation box is non-periodic on the top and bottom sides where the voltage leads are attached and the other sides of the box are kept periodic. The static external electric field is obtained by solving Poisson equation with appropriate boundary condition on the three dimensional spatial grid points of resolution $\sigma$.  The long range electrostatic Coulomb interactions among all the charged beads  are calculated exactly using the Ewald summation method~\cite{Allen} in 3D.
\par
\par
{\bf $\bullet$ Results:}~
We assume that the dsDNA is captured at the pore entrance and study the role of co- and counter-ions in the translocation process. Before we proceed to study translocation, we first verify and reproduce the well established results of counter-ion condensation on the same system studied by Jeon {\em et al.} ~\cite{Dobrynin2007} consisting of 128 bead long dsDNA of diameter $\sigma_{bead}=1.0$ with every third bead carrying a charge $q_{bead}=-1.0$. This corresponds to the charge fraction of the DNA backbone $f=1/3$. To maintain the charge neutrality, 42 counter-ions each having a charge $q^{+ve}_{ion}=1.0$ are added into the simulation.
The simulation box is cubic with length $L=107.4\sigma$ so that the ionic concentration is $10^{-4}/\sigma^3$. The polymer with counter-ions are equilibrated for 5 Rouse relaxation steps and Brownian dynamics simulations are carried out for different values of Bjerrum length ($l_B$).
Fig.~\ref{Model}(b) shows that after an initial increase the radius of gyration decreases as a function of $l_B$ and DNA polymer collapses from extended to a globular state confirming the previous findings of counter-ion condensation~\cite{Manning,Dobrynin2007,Stevens}.
\par
{\bf $\bullet$ Current Blockades in a Nanopore:}~
An external electric field causes a steady and delectable current flow through the pore. The origin of the current blockade in the nanopore can be attributed to several factors such as, physical obstructions, electrostatic interactions between the charged species, 
the nanopore geometry, ionic conductivity, and electric double layer formation are some of the important ones to mention.
The physical obstruction created by the effective volume fraction of the translocating species restricts the flow of ions through the nanopore.
When no obstacle is present, we refer to the constant flow of ionic current as the ``open pore current": $I_{free}$.
The presence of charged and bulkier species alter the local conductivity and resistance of the nanopore as they approach the pore orifice~\cite{Aksimentiev2019}. We assume that the current blockade is proportional to the volume of the species inside the pore, propose an ansatz for the volumetric current and justify {\em a posteori} by comparing the reconstructed current using this ansatz with the exact current through the pore. Validation of this ansatz will allow to simulate larger systems as discussed in the conclusion section.
\par
{\bf $\bullet$ Volumetric current blockade ansatz:}~
The volume of a DNA monomer $v_{bead} = \frac{4}{3}\pi\left(\frac{\sigma_{bead}}{2}\right)^3 = \frac{4}{3}\pi\sigma^3$ for  $\sigma_{bead} = 2\sigma$. Likewise, the oligomer monomers have a volume of $v_{motif} = \frac{1}{6}\pi\sigma^3$. At a particular instant of time, the occupancy of the DNA segment and the fraction of the oligomers inside the pore contribute to the current attenuation and we propose the following ansatz for the approximate current
\begin{equation}
  I_{BD}^{approx}(t) = I_{free}\left( 1-\frac{V_{vdw}(t)}{\gamma V_{pore}} \right ).
  \label{vol_I}
\end{equation}
Here, $V_{vdw}(t) = \sum_{i=1}^{N_b} \alpha_i(t)v_{bead} + \sum_{i=1}^{N_t}  \beta_{i}(t)v_{motif} $ is the total volume fraction caused by $N_d$ and $N_t$ number of the DNA and oligomer monomers inside the pore of
volume $V_{pore} = \pi r_{pore}^2 L_{pore} = 16 \pi \sigma^3$
causing the current blockade. Here $\alpha_i(t)$, $\beta_i(t)$, and $\gamma$ are adjustable parameters of the order of unity to fit the exact current blockade data. Experimentally the nanopores can be charged causing an electric double layer formation which will reduce the effective pore volume.  This can be incorporated by adjusting the parameter $\gamma \simeq1$.
\par
{\bf $\bullet$ Exact current blockade:}~
We construct the exact current traces by recording the co-ion and counter-ion flow through the nanopore as a function of the applied electric field. When no dsDNA segment is present inside the pore we observe a fluctuating number ($\sim 1-3$ per unit MD time) of both co-ions and counter-ions traveling across the nanopore. 
\begin{figure}[ht!]
\begin{center}
\includegraphics[width=0.48\textwidth]{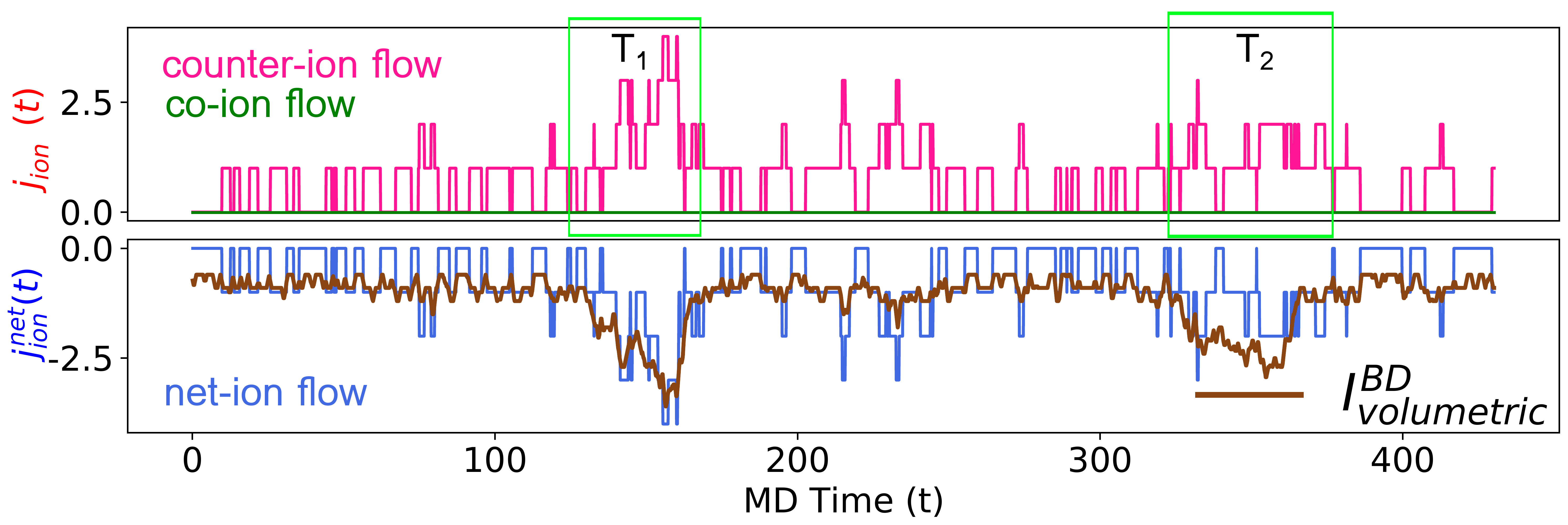}
\end{center}
\caption{\small (top) Counter-ion (Magenta) and co-ion (Green)  flow and through the nanopore during the translocation where the dwell time of $T_1$ and $T_2$ are indicated by the green rectangles. (bottom) The net ion flow (Cyan) and the scaled volumetric current blockade (Violet) where the volumetric current amplitude is scaled to match the peaks of the actual ion flow. The fitted current is less noisy.}
\label{Current}
\end{figure}
\par
However,  when a dsDNA segment (with or without the sidechain) enters into the nanopore it attracts the counter-ions resulting in a significant flow of counter-ions and almost negligible flow of co-ion flow as seen in our simulation snapshots (see Fig.~\ref{Model}(c)-(h)).  Thus, \textit{the net current
\begin{equation}
  j_{ion}^{net}= j_{ion}^{-}-j_{ion}^{+} \approx -j_{ion}^{+}
  \label{j_ion}
  \end{equation}
  (shown in Blue) approximately is a mirror image of the counter-ion current (shown in Magenta)}  in  Fig.~\ref{Current}. In Fig.~\ref{Current}(b) we also overlay the current blockade obtained from the volumetric occupation model (in dark maroon), validating the ansatz (Eqn.~\ref{vol_I}), albeit with less noise.
\par 
It is worth noting  that despite the complete attenuation of the co-ion flow, the net current through the pore is caused by the counter-ions drawn by the translocating dsDNA and the sidechain motifs. This argument is substantiated  by the observed upsurge in the counter-ion flow during $t=120-169$ and $t=323-371$ during the passage of the sidechain motifs. To confirm this origin of current in the nanopore, we further calculate the fraction of adsorbed counter-ions under mono-valent and di-valent salt concentrations discussed below.
\par
\begin{figure}[ht!]
\begin{center}
\includegraphics[width=0.48\textwidth]{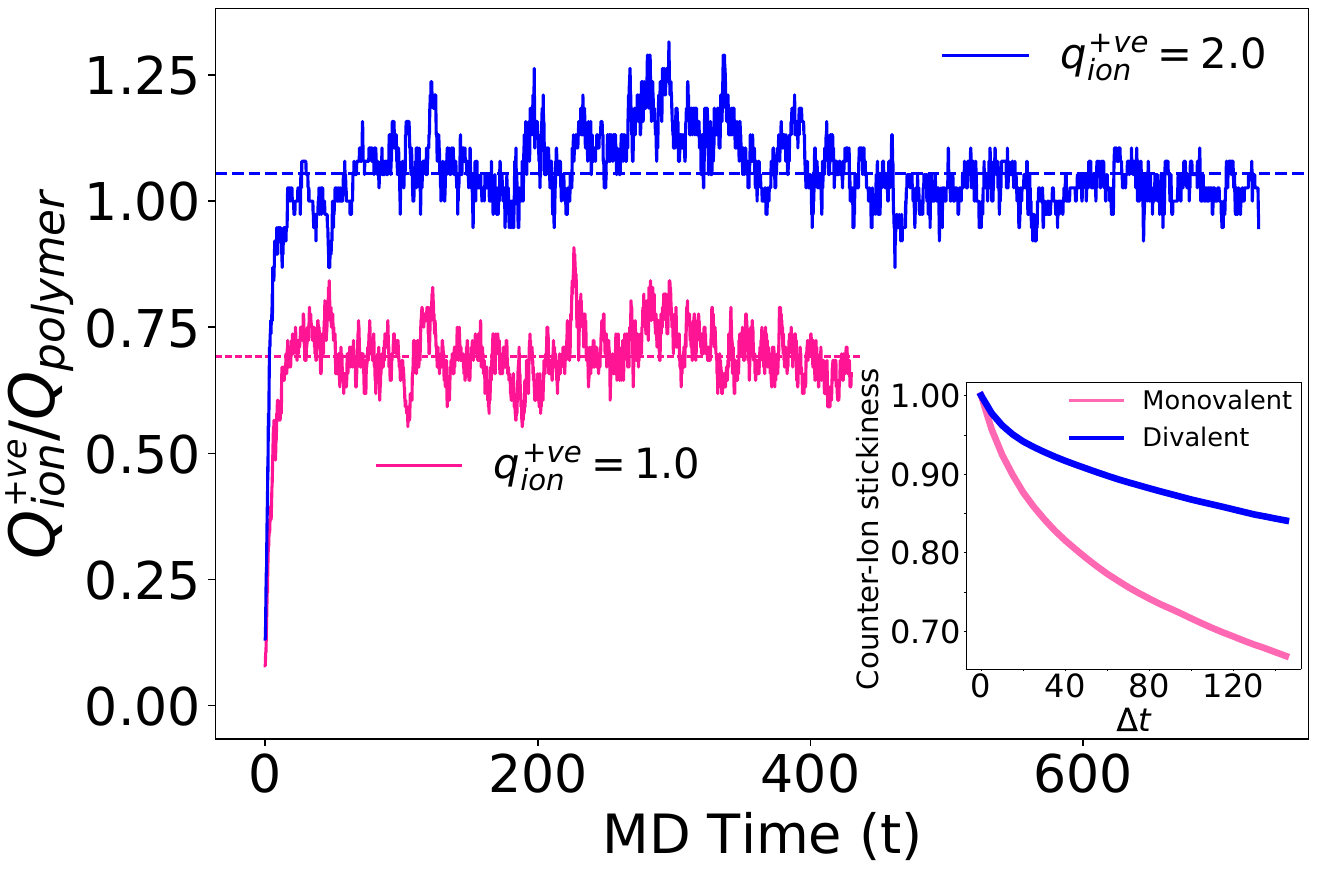}
\end{center}
\caption{\small The fraction $Q_{ion}/Q_{polymer}$ condensation of $Q_{ion}$ on the negatively charged dsDNA of net charge $Q_{polymer}$ as a function of translocation time in the presence of mono- (blue) and di-valent (Brown) salts. The inset represents the stickiness parameter that determines the residual correlation between the counter-ions near the dsDNA at different time intervals for the mono (Pink) /di-valent (Blue) salt. }
\label{ion_fraction}
\end{figure}

{\bf $\bullet$ Counter-Ion condensation for mono and di-valent salt condition:}~
The counter-ion condensation on a negatively charged DNA is well established phenomena and first studied by Manning {\em et al.}~\cite{Manning}. During the simulation 
 we kept track of  the fraction of counter-ions within a radius of 3~nm of the DNA backbone  condensing on the dsDNA. Fig.~\ref{ion_fraction} shows that a 10 fold increase of counter-ions near the
DNA backbone than the co-ions. It is interesting to note that under the conditions of divalent salt, the condensation of net counter-ion charge on the dsDNA backbone is higher, while the population of co-ions remains almost identical for both types of salts. We calculated the counter-ion stickiness parameter by determining the residual
correlation between the counter-ions present near the 3 nm radius of the dsDNA at different time intervals. The inset of Figure~\ref{ion_fraction} demonstrates that counter-ions from the divalent salt have a higher tendency to remain close to the DNA backbone. As a result, the counter-ion correlation function decays at a slower rate compared to the monovalent salt.\par
{\bf $\bullet$ Effect of counter-ion valence on translocation and dwell time:}~Several experimental studies have shown that salt containing ions of  higher valencies can increase the translocation time~\cite{Uplinger}. In a recent computational study, Hsiao {\em et al.} showed that trivalent ions slow down the translocation speed~\cite{Hsiao}. To further explore the effect of ion valencies, we simulated the same system only replacing the total number of monovalent ions (each of charge $q$) with half the number of divalent (each of charge $2q$) ions, 
\begin{figure}[ht!]
\begin{center}
\includegraphics[width=0.48\textwidth]{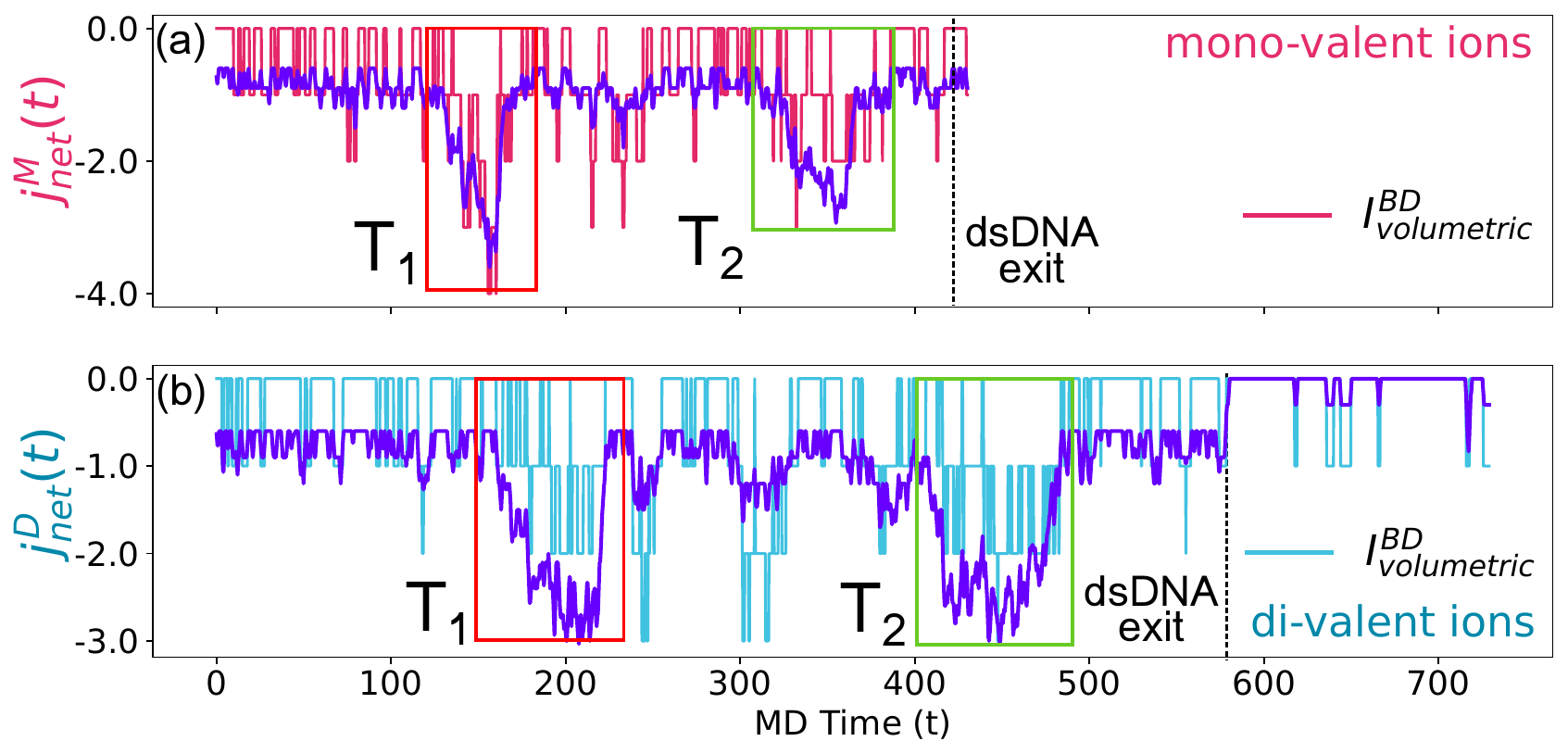}
\end{center}
   \caption{\small Comparison of net ion flow in the case of (a) mono-valent (same as Fig.~\ref{Current}(b)) and (b) di-valent salt. The pink/cyan lines show the net ion flow from the explicit consideration of co-ions and counter-ions in the BD simulation for mono/di-valent salts, and the violet lines are obtained from the volumetric occupation model. The ion flow associated with the passage of the sidechain motifs $T_1$ and $T_2$ are highlighted using red and green box respectively.}
\label{Ion-Flow_comparison}
\end{figure}
 maintaining electrical neutrality (see Table-I in supplementary materials). We observe that the in presence of di-valent ions the translocation time increases by roughly 1.5 fold shown in Fig.~\ref{Ion-Flow_comparison}(b).
\begin{figure}[ht!]
\begin{center}
\includegraphics[width=0.48\textwidth]{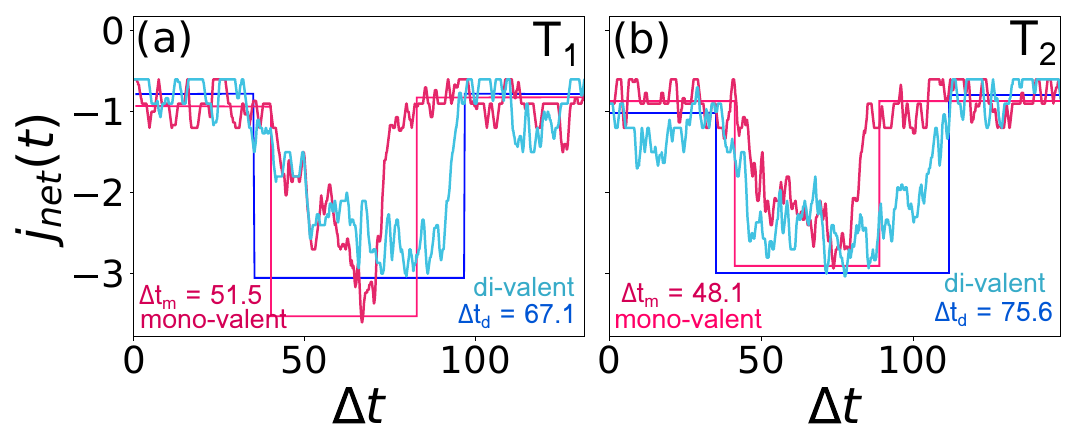}
\end{center}
\caption{\small Dwell time of (a) $T_1$ and (b) $T_2$ in presence of mono-valent (cyan) and di-valent (purple) ions. Two square waves are used to make an approximate fit to the current amplitude of the sidechain dwell-time denoted by $\Delta t_m$ and $\Delta t_d$ for mono-, and di-valent ions, respectively.}
\label{Dwell-time}
\end{figure} 
This in turn makes the sidechain dwell times pronounced and longer. Fig.~\ref{Dwell-time}(a)-(b) show the comparison of the blockade durations of $T_1$ and $T_2$, by approximating the volumetric current with a square wave, with the exact dwell times of the motifs. For both the motifs we notice that the dwell time increases by $\frac{\Delta t_m}{\Delta t_d}$, corresponds to a factor of 1.3 to 1.6. This shows that a di-valent salt solution makes the dwell time longer and therefore, will have a greater chance for the sidechain motifs and other features residing on a genome getting detected.\par
{\bf $\bullet$ Conclusions and outlook:}~
Many mesoscopic models to study DNA translocation through a nanopore use ordinary Brownian dynamics where the co-ions and counter-ions are not included for faster computation speed. These models have been successful in qualitatively reproducing experimental  results, such as, dwell time for a long dsDNA using tension propagation theory for longer polymer chains. Evidently, by construction, a direct calculation of current blockade is not possible with a bead-spring model of a polymer using the BD scheme. In this manuscript we extended the BD scheme to an EKBD scheme by putting co-ions and counter-ions interacting via long-range Coulomb interaction with each other and with the translocating DNA, enabling us to calculate the current blockade characteristics exactly. An important and less obvious of our findings is that the drop in the conventional (positive) current through the pore is due to the condensation of the counter-ions on the translocating DNA and not so much due to drop in the co-ions passing through the pore. This finding aligns with previous studies conducted by Holm {\em et al.}~\cite{Holm}, Tanaka {\em et al.}~\cite{Rabin}, and  Cui~\cite{Cui2010}. This counter-intuitive behavior challenges the conventional understanding of ion movement in nanopore systems under low salt concentration and sheds light on the complex interplay between DNA-motifs and ions during translocation. However, it should be noted that at higher salt concentrations, the electrostatic screening effect becomes more prominent, and the exclusion volume effect becomes the primary cause of the current blockade. This condensation increases for the divalent counter-ions resulting a slowing down of the entire translocation process. \par
Another motivation for this more expensive EKBD study is to reconstruct the exact current blockade for the DNA bound  molecular features using a volumetric ansatz. We find that the reconstructed current compares well with the exact current blockade. Because we integrate out the faster degree of freedom of the co-ions and counter-ions the added bonus is that this reconstructed current is less noisy. This constitutes the 1st step of reduction to reconstruct the current blockade.  In the second step we repeat the same ansatz using a BD scheme (without the co-ions and the counter-ions) with a rescaling of the effective volume to account for the absence of the co-ions and counter-ions. A comparison of the two volumetric ansatz will allow to reconstruct the actual current blockade. Since BD scheme without the co-ions and counter-ions is several orders of magnitude faster this mapping procedure will allow reconstruction of the current traces for ultra long genomic reads. This procedure will also provide insights how to map the units of time scales with and without the co- and counter-ions and will help to construct a theory for such mapping. Previously we and others used a similar criteria using P\'{e}clet number~\cite{Seth_DNP,deHaan,Stein}. We studied an ideal system without charges residing at the surface of the nanopore. It is possible to study the effect of possible double layer formation and its effect on the translocating DNA with this EKBD scheme. By incorporating explicit solvent molecules (such as, TIP3P model of water) it will also be possible to assess the relative role of electrokinetic and hydrodynamic effects which again will provide useful information to make a unified theoretical scheme for nanopore translocation.\par
{\bf $\bullet$ Acknowledgments:}~
This research has been supported by a subcontract by The Royal Institution for the Advancement of Learning, McGill, University, Canada, from the parent grant 1R21HG011236-01 from the National Human Genome Research Institute at the National Institute of Health.
All computations were carried out using STOKES High Performance Computing Cluster at UCF. The simulation movies were generated using VMD software package.
\vfill

\end{document}